\documentclass[journal=jctcce,manuscript=article,layout=traditional]{achemso}
\usepackage{chemformula} 
\usepackage[T1]{fontenc} 
\usepackage{caption}
\usepackage[utf8]{inputenc}
\usepackage{graphicx}
\usepackage{tabularx}
\usepackage{amsmath}
\usepackage{siunitx}
\usepackage{physics}
\usepackage{textgreek}
\usepackage{xfrac}
\usepackage{xcolor}
\usepackage{bm}
\usepackage{natbib}
\usepackage{bbold}
\usepackage{hyperref}
\usepackage{multicol}
\usepackage{appendix}
\usepackage[capitalise]{cleveref} 
\DeclareSIUnit\angstrom{\protect \text {Å}}

\author{Timothy N. Georges}
\affiliation[University of Oxford]{Department of Chemistry, Physical and Theoretical Chemistry Laboratory,\\ University of Oxford, Oxford OX1 3QZ, UK}
\alsoaffiliation[Balliol College]{Balliol College, University of Oxford, Oxford OX1 3BJ, UK}
\email{timothy.georges@chem.ox.ac.uk}

\author{Louis Summerley}
\affiliation[University of Oxford]{Department of Chemistry, Physical and Theoretical Chemistry Laboratory,\\ University of Oxford, Oxford OX1 3QZ, UK}
\alsoaffiliation[Worcester College]{Worcester College, University of Oxford, Oxford OX1 2HB, UK}

\author{Johan E. Runeson}
\affiliation[University of Freiburg]{Institute of Physics, University of Freiburg, Freiburg 79104, Germany}

\author{\\William Barford}
\affiliation[University of Oxford]{Department of Chemistry, Physical and Theoretical Chemistry Laboratory,\\ University of Oxford, Oxford OX1 3QZ, UK}
\alsoaffiliation[Balliol College]{Balliol College, University of Oxford, Oxford OX1 3BJ, UK}
\email{william.barford@chem.ox.ac.uk}

\title[A Vibronic Coupling Model to Study the Nonadiabatic Dynamics of Polyenes]
  {A Vibronic Coupling Model to Study the Nonadiabatic Dynamics of Polyenes}

\begin{document}







\begin{abstract}
We develop a linear vibronic coupling (LVC) model for polyenes described by the extended Hubbard-Peierls Hamiltonian. This model is applied to \textit{trans}-hexatriene to benchmark quantum-classical dynamics methods against fully quantum simulations. We find that surface-hopping methods describe short times more accurately than multi-trajectory Ehrenfest. None of the quantum-classical methods studied obtain the long-time population oscillations found in fully quantum simulations. Varying the parameters of the LVC Hamiltonian, we find that surface hopping reproduces the correct trends in the long-time dynamics across a wide range of parameters, but generally overestimates the degree of internal conversion. On the other hand, multi-trajectory Ehrenfest gives more accurate long-time populations in proximity to the hexatriene parameter set. 
\end{abstract}

\section{Introduction}\label{sec:intro}

Polyene derivatives have attracted interest in recent decades owing to the observation of singlet fission, a process where a singlet excited state decays into two triplet states\cite{smithSingletFission2010}. For example, carotenoids have been reported exhibiting singlet fission both as aggregates\cite{wangHighYieldSingletFission2010,musserNatureSingletExciton2015,kunduPhotogenerationLongLivedTriplet2021,pengInvestigationUltrafastIntermediate2024} and in a biological environment\cite{yuCarotenoidSingletFission2017}. Moreover, singlet fission has been observed in diphenylhexatriene and its derivatives\cite{huangCompetitionTripletPair2021,millingtonSynthesisIntramolecularSinglet2023,millingtonSolubleDiphenylhexatrieneDimers2023}. A secondary absorber in a solar cell that absorbs at the correct wavelength and is capable of singlet fission could provide a mechanism for solar cells to exceed the Schockley-Queisser limit\cite{shockleyDetailedBalanceLimit1960,hannaSolarConversionEfficiency2006}. 

The mechanism for singlet fission in carotenoid derivatives is still poorly understood, with two competing hypotheses. One hypothesis is that singlet fission occurs from an intramolecular triplet-pair state\cite{wangLowlyingSingletStates2005,polivkaDarkExcitedStates2009,valentineHigherenergyTripletpairStates2020,kunduPhotogenerationLongLivedTriplet2021,barfordTheoryDarkState2022,manawaduSingletTripletPairProduction2022,manawaduPhotoexcitedStateDynamics2023,barford_singlet_2023,barford_theory_2023,ichertSingletFissionCarotenoid2025}, while the other invokes bimolecular charge-transfer states\cite{casanova_theoretical_2018,pengInvestigationUltrafastIntermediate2024}, both of which are populated via the photoexcited state. Competing with either mechanism is internal conversion to the lowest energy singlet triplet-pair state, i.e., the $2A_g$ state.

To test these hypotheses, \textit{ab initio} methods are generally too expensive to calculate the multiple potential energy surfaces required for dynamics\cite{taffetStateFallsCarotenoidrelevant2018,taffetCarotenoidNuclearReorganization2019,khokhlovInitioStudyLowLying2020,chagasLowlyingExcitedStates2025}. Model Hamiltonians, on the other hand, require much lower computational cost to solve large carotenoid systems. They include a limited number of interactions, allowing one to understand the minimum interactions to qualitatively describe a process. An example is the extended Hubbard-Peierls Hamiltonian of $\pi$-electron systems, which Manawadu and coworkers have simulated using the time-dependent density matrix renormalization group (TD-DMRG) method with single-trajectory Ehrenfest dynamics\cite{manawaduSingletTripletPairProduction2022,manawaduPhotoexcitedStateDynamics2023}. The extended Hubbard-Peierls Hamiltonian was solved on-the-fly and one Ehrenfest trajectory was performed, starting in the relaxed geometry of the ground state. Because of the quantum-classical forces used in Ehrenfest, there is no coupling between states in different symmetry sectors and thus this approach cannot describe vibronic transitions between the $B_u$ and $A_g$ electronic manifolds for a system with $C_{2h}$ symmetry\cite{manawaduSingletTripletPairProduction2022}. This method can only describe transitions between the $B_u$ and $A_g$ electronic manifolds by explicitly breaking $C_2$ symmetry\cite{manawaduPhotoexcitedStateDynamics2023}. For carotenoids, each trajectory is too computationally expensive for multi-trajectory methods.

Our future goal is to model the excited state dynamics of lycopene, a carotenoid with $C_{2\text{h}}$ symmetry that undergoes singlet fission in aggregates\cite{kunduPhotogenerationLongLivedTriplet2021,pengInvestigationUltrafastIntermediate2024,magneSingletFissionHeterogeneous2025}. We aim to predict the timescales for the competing processes of singlet fission and internal conversion to the $2A_g$ state. Therefore, we require a method that describes nonadiabatic dynamics accurately for up to $22$ conjugated carbon atoms. The method must also be able to simulate more than two states, because for longer polyenes other triplet-pair states have similar energy to the $1B_u$ and $2A_g$ states. Finally, we need to describe transitions between the $B_u$ and $A_g$ electronic manifolds and thus require an approach that goes beyond the Born-Oppenheimer approximation.

Fully quantum \textit{ab initio} methods are untenable for carotenoids, because as the polyene length increases the Hilbert space increases exponentially. Thus, to account for all of the criteria listed above, to model lycopene we will use a vibronic coupling Hamiltonian\cite{koppelMultimodeMolecularDynamics1984} constructed via the extended Hubbard-Peierls Hamiltonian. Since the electronic states of polyenes are highly correlated\cite{BarfordBook}, we will calculate the potential energy surfaces and inter-state coupling constants using the density matrix renormalisation group (DMRG) method\cite{whiteDensityMatrixFormulation1992,whiteDensitymatrixAlgorithmsQuantum1993,manawaduDynamicalSimulationsCarotenoid2023}.

The goals of the present paper are two-fold. First, we construct a linear vibronic coupling (LVC) model for polyenes described by the extended Hubbard-Peierls Hamiltonian. Second, using this LVC model we benchmark various quantum-classical nonadiabatic dynamics methods against fully quantum simulations. We take hexatriene as our model system, as this has already been studied theoretically\cite{woywodVibronicCouplingTrans2000a,woywodVibronicCouplingTrans2000,komaindaInitioQuantumStudy2015,komaindaInitioBenchmarkStudy2016}. Komainda and coworkers constructed vibronic coupling models of \textit{trans}- and \textit{cis}-hexatriene, parameterized by \textit{ab initio} potential energy surfaces along selected internal coordinates\cite{komaindaInitioQuantumStudy2015,komaindaInitioBenchmarkStudy2016}, and simulated their quantum dynamics with the multi-configurational time-dependent Hartree (MCTDH) method\cite{beckMulticonfigurationTimedependentHartree2000}. They showed internal conversion from $1B_u$ to $2A_g$ and absorption spectra with some qualitative agreement to experiment\cite{andersonInternalConversionsTrans2000,komaindaInitioBenchmarkStudy2016}. However, their \textit{ab initio} potential is not feasible for longer polyenes, whereas the extended Hubbard-Peierls Hamiltonian is. We show that our protocol predicts diabatic populations in agreement with those of Ref.~\citenum{komaindaInitioBenchmarkStudy2016}. 

There exists a plethora of viable nonadiabatic dynamics methods and there is ongoing discussion in the field on how to choose between and correctly benchmark them\cite{ibeleMolecularPerspectiveTully2020,gomezBenchmarkingNonadiabaticQuantum2024,cigrangRoadmapMolecularBenchmarks2025}. In order to choose
a suitable method for carotenoids, we benchmark three quantum-classical approaches: Multi-trajectory Ehrenfest (MTE), a simple mean-field approach; fewest-switches surface hopping (FSSH), which is well-established for the nonadiabatic dynamics of conjugated polymers, organic semiconductors and vibronic coupling models\cite{tully_molecular_1990,nelsonNonadiabaticExcitedStateMolecular2014,xiePerformanceMixedQuantumClassical2020,roostaEfficientSurfaceHopping2022,plasser_highly_2019,zobel_surface_2021}; and the multi-state mapping approach to surface hopping (MASH) as introduced by Runeson and Manolopoulos\cite{mannouch_mapping_2023,runeson_multi-state_2023}, which has been applied to ultrafast dynamics in cyclobutanone,\cite{huttonUsingMultistateMapping2024}  light-harvesting complexes\cite{runesonNuclearQuantumEffects2025}, and charge transport in organic semiconductors\cite{runesonChargeTransportOrganic2024}. We investigate which method is best suited for carotenoid simulations by comparing against fully quantum results found with the short iterative Lanczos propagator (SILP) method\cite{domckeConicalIntersectionsElectronic2004}. The observables we aim to reproduce with quantum-classical methods are the diabatic state populations and the nuclear positions. 

The outline of this paper is as follows.~\cref{sec:theory} introduces the extended Hubbard-Peierls model and describes our derivation of a linear vibronic coupling model from it.~\cref{sec:methods} describes the four quantum dynamics methods that we apply to our LVC model, while~\cref{sec:results} describes our results. As well as comparing methods at the realistic parameter set, we also investigate the reliability of the quantum-classical methods over a range of model parameters. Finally, we conclude in~\cref{sec:conclusion} and discuss forthcoming work.

\section{Theory}\label{sec:theory}
In this Section, we introduce our approach to construct a linear vibronic coupling Hamiltonian for polyenes. First, we describe the electronic structure method used to calculate adiabatic potential energy surfaces, the extended Hubbard-Peierls Hamiltonian. Then, we define our coordinate system and the LVC Hamiltonian. Finally, we discuss how we calculate the parameters for the LVC Hamiltonian from the extended Hubbard-Peierls Hamiltonian using exact diagonalization.

\subsection{Extended Hubbard-Peierls Hamiltonian}
In this work, we compute the electronic structure of polyenes with the extended Hubbard-Peierls Hamiltonian. We start with the UV-Peierls (UVP) Hamiltonian that includes electron-electron interactions, electron kinetic energy and electron-nuclear coupling:
\begin{equation}\label{eq:H_UVP_elec}
    \hat{H}_{\text{UVP}}=U\sum_{n=1}^N\left(\hat{N}_{n\uparrow}-\frac{1}{2}\right)\left(\hat{N}_{n\downarrow}-\frac{1}{2}\right)+V\sum_{n=1}^{N-1}\left(\hat{N}_{n+1}-1\right)\left(\hat{N}_{n}-1\right)-2\sum_{n=1}^{N-1}t_n\hat{T}_n,
\end{equation}
where $N$ is the number of carbon atoms and $U$ and $V$ are the on-site and nearest-neighbor Coulomb repulsion terms, respectively. The third term combines electron kinetic energy and electron-nuclear interactions, in which $t_n$ are the bond hopping integrals given by 
\begin{equation}\label{eq:tn_defn}
    t_n=t_0-\alpha\left(x_{n+1}-x_n\right),
\end{equation}
where $x_n$ are the nuclear positions, $\alpha$ is the electron-nuclear coupling constant and $t_0$ is the average bond hopping integral. The electron number operator is
\begin{equation}\label{eq:Nn_defn}
    \hat{N}_n=\sum_{\sigma}\hat{N}_{n\sigma}=\sum_{\sigma}\hat{c}^{\dagger}_{n\sigma}\hat{c}_{n\sigma},
\end{equation}
where $\hat{c}^{\dagger}_{n\sigma}$ creates an electron in the $n^{\text{th}}$ carbon $2p_z$ orbital with spin $\sigma$. The bond hopping operator is
\begin{equation}
    \hat{T}_n=\frac{1}{2}\sum_{\sigma}\left(\hat{c}^{\dagger}_{n+1\sigma}\hat{c}_{n\sigma}+\hat{c}^{\dagger}_{n\sigma}\hat{c}_{n+1\sigma}\right).
\end{equation}
We also add a term to the Hamiltonian that breaks particle-hole symmetry, but maintains inversion symmetry,
\begin{equation}
    \hat{H}_{\text{SB}}=\sum_{n=1}^N\epsilon_n\left(\hat{N}_n-1\right).
\end{equation}
The values of $\epsilon_n$ are listed in the Supporting Information.  For a given molecule, the particle-hole symmetry-breaking term is found by matching the carbon electron densities of the extended Hubbard-Peierls Hamiltonian to the Mulliken charge densities of a DFT calculation, using a gradient descent algorithm. We use the B3LYP functional and def2-TZVP basis set, following the procedure set out by Manawadu and coworkers\cite{manawaduPhotoexcitedStateDynamics2023}.
Since we can no longer specify particle-hole symmetry, our polyene states have term symbols $A_g$ or $B_u$. $A_g$ states are symmetric with respect to a $C_2$ rotation, while $B_u$ states are anti-symmetric. Molecular vibrations with $B_u$ symmetry cause transitions between electronic states in these two sectors.

The elastic nuclear energy is
\begin{equation}\label{eq:H_ph}
    \hat{H}_{\text{C-ph}}=\frac{K}{2}\sum_{n=1}^{N-1}\left(x_{n+1}-x_n\right)^2
    -2\alpha\Gamma\sum_{n=1}^{N-1}\left(x_{n+1}-x_n\right),
\end{equation}
where $K$ is the force constant of a carbon-carbon $\sigma$ bond\cite{ehrenfreundAmplitudePhaseModes1987}. The final term penalizes deviation from a fixed chain length. $\Gamma$ is the mean of the expectation value of the bond order operator in the ground state\cite{BarfordBook},
\begin{equation}\label{eq:Gamma_defn}
    \Gamma=\frac{1}{N-1}\sum_{n=1}^{N-1}\mel{1A_g}{\hat{T}_n}{1A_g},
\end{equation}
and is found iteratively to be $\Gamma=1.3$.

The extended Hubbard-Peierls Hamiltonian used to calculate the Born-Oppenheimer potential energy surfaces is 
\begin{equation}\label{eq:ext_hubb_ham}
    \hat{H}_{\text{Ext-Hub}}=\hat{H}_{\text{UVP}}+\hat{H}_{\text{SB}}+\hat{H}_{\text{C-ph}}.
\end{equation}
To be consistent with our previous work on carotenoids, we choose the following parameters: $U=7.25~\unit{\eV}$, $t_0=2.40~\unit{\eV}$, $\alpha=4.593~\unit{\eV \angstrom^{-1}}$, $K=46~\unit{\eV\angstrom^{-2}}$ and $m=K/\omega_0^2$, where $\omega_0=2.15\times 10^{14}~\unit{\s^{-1}}$ \cite{ehrenfreundAmplitudePhaseModes1987,barfordDensitymatrixRenormalizationgroupCalculations2001,manawaduPhotoexcitedStateDynamics2023}. These parameters were chosen to accurately predict excited state energies of long polyenes in the condensed phase.

It is difficult to calculate the relative energy of the low-lying excited states of polyenes with \textit{ab initio} methods, because of the contrasting character of the covalent $2A_g$ and ionic $1B_u$ states\cite{chagasLowlyingExcitedStates2025}. Park and coworkers found vertical energy gaps of about $0.7~\unit{\eV}$ with $\delta$-CR-EOMCC(2,3), $0.5~\unit{\eV}$ with XMSCASPT2 and $0.4~\unit{\eV}$ with XMCQDPT2\cite{parkInternalConversionBright2021}. Chagas and coworkers reported vertical energy gaps of  $0.26~\unit{eV}$ with CASPT2 and $0.22~\unit{\eV}$ with MR-CISD+P\cite{chagasLowlyingExcitedStates2025}. Komainda and coworkers found a vertical energy gap of $0.29~\unit{\eV}$ with MSCASPT2 and $0.37~\unit{\eV}$ with DFT/MRCI\cite{komaindaInitioBenchmarkStudy2016}. In our extended Hubbard-Peierls Hamiltonian,~\cref{eq:H_UVP_elec}, we use the parameter $V=3.35~\unit{\eV}$ so that the vertical energy gap between the $1B_u$ and $2A_g$ states of hexatriene is $0.29~\unit{\eV}$, as found in Ref.~\citenum{komaindaInitioBenchmarkStudy2016}. This value ensures ultrafast energy crossover of the states on the optical mode. 

\subsection{Vibronic Coupling Hamiltonian}\label{sec:vibronic_coupling}
\begin{figure}
    \centering
    \includegraphics[]{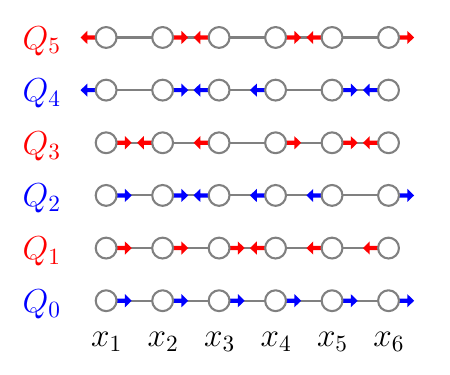}
    \caption{A schematic diagram of the normal coordinates of the ground state of the extended Hubbard-Peierls Hamiltonian, ordered by increasing energy. Each circle represents a carbon atom of hexatriene. Symmetric ($A_g$) modes are red and anti-symmetric ($B_u$) modes are blue. The lowest-energy mode, $Q_0$, is translational, so it is excluded from the LVC model.}
    \label{fig:normal_coords}
\end{figure}

We construct a vibronic coupling Hamiltonian in terms of the normal coordinates of the one-dimensional extended Hubbard-Peierls Hamiltonian. The normal coordinate transformation is
\begin{equation}
    \mathbf{x}=\mathbf{Bq},
\end{equation}
where the matrix $\mathbf{B}$ satisfies $\mathbf{B^TB}=\mathbb{1}$ and $\mathbf{B^T}\mathcal{H}^{(x)}\mathbf{B}=\mathbf{K}^{(0)}$. $\mathcal{H}^{(x)}$ is the Hessian of the ground state of the extended Hubbard-Peierls Hamiltonian and $\mathbf{K}^{(0)}$ is the diagonal matrix of the force constants of the normal modes. The columns of $\mathbf{B}$ are the normal coordinates in terms of the Cartesian coordinates $\mathbf{x}$. These coordinates are illustrated in~\cref{fig:normal_coords}. The coordinate $Q_0$ corresponds to pure translation and is removed. In terms of internal coordinates, the Hessian is
\begin{equation}\label{eq:internal_hessian}
    \mathcal{H}^{(r)}=\frac{\partial^2}{\partial r_i\partial r_j} \mel{\text{GS}}{\hat{H}_{\text{Ext-Hub}}}{\text{GS}}
    = 2\alpha \frac{\partial}{\partial r_i}\mel{\text{GS}}{\hat{T}_j}{\text{GS}} + K\delta_{ij}.
\end{equation}
The internal coordinates $r_i$ are related to Cartesian coordinates by $r_n=(x_n-x_{n+1})$, such that $\mathbf{r}=\mathbf{Ax}$. The Hessian in terms of atomic displacements is
\begin{equation}\label{eq:cartesian_hessian}
    \mathcal{H}^{(x)}=\mathbf{A^T}\mathcal{H}^{(r)}\mathbf{A}.
\end{equation}
The second term on the right-hand side of~\cref{eq:internal_hessian} is the Hessian of coupled one-dimensional harmonic oscillators and the first term on the right-hand side is approximated via the central difference theorem,
\begin{equation}
    f^{\prime}(x_0)\approx\frac{f(x_0+h)-f(x_0-h)}{2h}+\mathcal{O}(h^2),
\end{equation}
where $h$ is a small displacement. In practice, $h$ is limited by the accuracy to which we can calculate the expectation value of the bond order operator. We used $h=1.0\times 10^{-5}\unit{\angstrom}$.

The two-state vibronic coupling Hamiltonian is\cite{cederbaumMultimodeVibronicCoupling1981,koppelMultimodeMolecularDynamics1984}
\begin{equation}\label{eq:H_2x2_vibronic}
    \hat{H}_{\text{C-VC}} = T_N\mathbb{1} + 
    \begin{pmatrix}
        W_{11}(\mathbf{q}) & W_{12}(\mathbf{q}) \\
        W_{21}(\mathbf{q}) & W_{22}(\mathbf{q}) \\
    \end{pmatrix},
\end{equation}
where the kinetic energy $T_N$ is
\begin{equation}
    T_N=\frac{1}{2m}\mathbf{p^Tp}.
\end{equation}
For normal-mode coordinates, $\mathbf{p}=m\dot{\mathbf{q}}$ is the momentum conjugate to $\mathbf{q}$. The first electronic state is $1B_u$ and the second electronic state is $2A_g$. For a linear vibronic coupling model, we have
\begin{equation}
    W_{ii}(\mathbf{q})=E^{(i)}+\sum_{g}\kappa^{(i)}_gq_g+\frac{1}{2}\sum_{\alpha}K_{\alpha}^{(0)}q_{\alpha}^2
\end{equation}
and
\begin{equation}
    W_{12}(\mathbf{q})=W_{21}(\mathbf{q})=\sum_u\lambda_uq_u.
\end{equation}
All sums are over normal modes, but a sum over $g$ includes only symmetric modes. Similarly, a sum over $u$ indicates a sum over only anti-symmetric modes. In the LVC model, each potential energy surface uses the force constant of the ground state, $K_{\alpha}^{(0)}=m\omega_{\alpha}^2$.

We introduce quantum phonons to the LVC Hamiltonian with quantum harmonic oscillator raising and lowering operators for each mode. The normal coordinates and their conjugate momenta are expressed in terms of quantum-phonon operators as follows:
\begin{equation}\label{eq:qi_qph}
    \hat{q}_{\alpha}=\left(\frac{\hbar}{2m\omega_{\alpha}}\right)^{\frac{1}{2}}\left(\hat{a}_{\alpha}^{\dagger}+\hat{a}_{\alpha}\right)
\end{equation}
and
\begin{equation}\label{eq:pi_qph}
    \hat{p}_{\alpha}=i\left(\frac{m\hbar\omega_{\alpha}}{2}\right)^{\frac{1}{2}}\left(\hat{a}_{\alpha}^{\dagger}-\hat{a}_{\alpha}\right).
\end{equation}
The dimensionless position operator $\hat{Q}_{\alpha}$ is
\begin{equation}\label{eq:Qalpha_defn}
    \hat{Q}_{\alpha}=\frac{1}{\sqrt{2}}\left(\hat{a}_{\alpha}^{\dagger}+\hat{a}_{\alpha}\right)=\frac{\hat{q}_{\alpha}}{l_{\alpha}},
\end{equation}
where for each mode the length scale $l_{\alpha}$ is 
\begin{equation}\label{eq:li_defn}
    l_{\alpha}=\left(\frac{\hbar}{m\omega_{\alpha}}\right)^{\frac{1}{2}}.
\end{equation}
This gives the quantum-phonon LVC Hamiltonian:
\begin{equation}\label{eq:h_quibronic}
    \hat{H}_{\text{Q-VC}} = \sum_{\alpha}\hbar\omega_{\alpha}\left(\hat{a}^{\dagger}_{\alpha}\hat{a}_{\alpha}+\frac{1}{2}\right)\mathbb{1} +
    \begin{pmatrix}
        E^{(1)}+\sum_{g}\kappa^{(1)}_gl_g \hat{Q}_{g} & \sum_{u}\lambda_{u}l_u \hat{Q}_u \\
        \sum_{u}\lambda_{u}l_u \hat{Q}_u  & E^{(2)}+\sum_{g}\kappa^{(2)}_gl_g \hat{Q}_{g}\\
    \end{pmatrix}.
\end{equation}
The linear intra- and inter-state coupling parameters in units of energy are $\Tilde{\kappa}^{(i)}_{\alpha}=l_{\alpha}\kappa^{(i)}_{\alpha}$ and $\Tilde{\lambda}^{(i)}_{\alpha}=l_{\alpha}\lambda^{(i)}_{\alpha}$. The parameters of the Hamiltonian are $E^{(1)} = 3.92795~\unit{\eV}$ and $E^{(2)} = 4.21704~\unit{\eV}$, with the others listed in~\cref{tab:lvc_parameters}.

\begin{table}[]
    \centering
    \begin{tabular}{|c|c|c|c|c|c|}
        \hline
        Mode, $\alpha$ & $1$ & $2$ & $3$ & $4$ & $5$ \\ 
        \hline 
        $\omega_\alpha$ & $0.07203$ & $0.13801$ & $0.19692$ & $0.22524$ & $0.24041$ \\ 
        \hline 
        $\Tilde{\kappa}^{(1)}_\alpha$ & $-0.06433$ & --- & $0.01340$ & --- & $-0.34202$ \\ 
        \hline 
        $\Tilde{\kappa}^{(2)}_\alpha$ & $-0.08505$ & ---  & $0.17861$ & --- & $-0.64464$ \\ 
        \hline 
        $\Tilde{\lambda}_\alpha$ & ---  & $ 0.03267$ & --- & $ 0.08692$ & --- \\ 
        \hline 
    \end{tabular}
    \caption{Parameters of the linear vibronic coupling Hamiltonian of hexatriene,~\cref{eq:h_quibronic}. The normal modes, $\alpha$, are illustrated in Fig.\ 1. The potential energy surfaces are calculated using the extended Hubbard-Peierls Hamiltonian. All values are in $\unit{\eV}$.}
    \label{tab:lvc_parameters}
\end{table}

\subsection{Potential Energy Surface Calculation and Fitting}
We use exact diagonalisation to calculate the adiabatic potential energy surfaces of the extended Hubbard-Peierls Hamiltonian, given in~\cref{eq:ext_hubb_ham}. Exact diagonalization is also used to find the polyene ground-state geometry. When calculating potential energy surfaces of the symmetric modes, $C_2$ symmetry is imposed to identify $A_g$ and $B_u$ states through crossings. 

The relaxed geometry of an eigenstate of the system is found by setting the force on each bond to zero and using the Hellmann-Feynman theorem\cite{BarfordBook},
\begin{equation}\label{eq:Hellmann-Feynman}
    f_n=-\frac{\partial\mel{\Psi}{\hat{H}}{\Psi}}{\partial r_n}=-\mel{\Psi}{\frac{\partial{\hat{H}}}{\partial r_n}}{\Psi}=0.
\end{equation}
Taking the derivative of the Hamiltonian $\hat{H}_{\text{Ext-Hub}}$ gives a self-consistent equation that is used to find the equilibrium bond hopping integrals,  
\begin{equation}\label{eq:tn_selfconsistent}
    t_n = t_0-\frac{2\alpha^2}{K}\left(\Gamma - \expval{\hat{T}_n}\right).
\end{equation}
We use this algorithm to calculate the ground-state geometry, such that the energy converges to $0.01~\unit{\eV}$.

All LVC Hamiltonian parameters in the diabatic basis are found by fitting to adiabatic potential energy surfaces. To calculate the adiabatic potential energy surfaces, we vary a single normal mode about the ground-state geometry. For symmetric modes, the parameters $\kappa_g^{(i)}$ are found by fitting a cut of the potential energy surface on a symmetric mode to a $15^{\text{th}}$ order polynomial using least squares fitting and taking the first order term. We find the linear inter-state coupling constants $\lambda_u$ by calculating the LVC Hamiltonian in the diabatic basis where all coordinates are zero, except a single anti-symmetric mode, $Q_u$. This Hamiltonian is diagonalized into the adiabatic basis for a range of values of $Q_u$, initially with a guess value of $\lambda_u$. Then, we calculate the sum of the squares of the difference between these approximate adiabatic potential energy surfaces and their respective adiabatic potential energy surfaces found from the extended Hubbard-Peierls Hamiltonian. The sum of squares is minimized using the Newton–Raphson method. All fits were done in the window $-0.5~\unit{\angstrom}$ to $+0.5~\unit{\angstrom}$.

\begin{figure}
    \centering
    \includegraphics[]{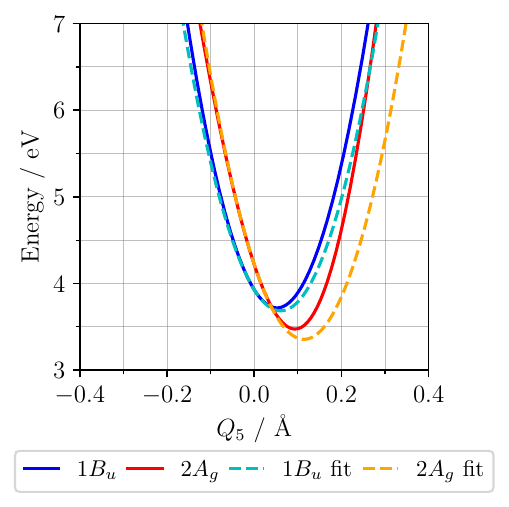}
    \caption{The cut of the potential energy surfaces of the two lowest-lying excited states along the highest energy symmetric mode, $Q_5$. The surfaces are calculated for the extended Hubbard-Peierls Hamiltonian. At the Franck-Condon point, the $1B_u$ lies $0.3~\unit{\eV}$ below the $2A_g$. However, increasing $Q_5$ results in the relaxation of the $2A_g$ below the $1B_u$, with the $2A_g$ minimum at about $0.1~\unit{\angstrom}$. The fit could be improved by including state-specific force constants and higher-order terms. However,~\cref{fig:hex_2state_q_comparison} shows that the poorly-fitted region is rarely entered during dynamics.}
    \label{fig:PES_Q5}
\end{figure}

\cref{fig:PES_Q5} shows the cuts of the potential energy surfaces of the two lowest-lying excited states along the highest energy symmetric mode, $Q_5$. This mode is the optical mode and is responsible for the relaxation of the $2A_g$ below the $1B_u$. The minima of the $1B_u$ and $2A_g$ states are $0.05~\unit{\angstrom}$ and $0.1~\unit{\angstrom}$ away from the ground-state geometry, respectively. The other symmetric, or tuning, modes do not result in the $2A_g$ relaxing below the $1B_u$. However, oscillation on these modes still changes the energy gap between the $2A_g$ and $1B_u$ states. The anti-symmetric, or coupling, modes are symmetric about $\mathbf{q}=0$ and oscillation on these modes linearly affects the coupling between the two electronic states. Cuts of the potential energy surfaces along the other four normal modes are shown in Figs.~S1 to S4 of the Supporting Information.

The LVC Hamiltonian poorly fits the minima of the excited states on the optical mode, as illustrated in~\cref{fig:PES_Q5}. In principle, this could be improved by including state-specific quadratic terms and higher-order polynomial terms. As shown in~\cref{sec:hexatriene}, oscillations on the $Q_5$ mode remain below $0.16~\unit{\angstrom}$ throughout dynamics and for all methods. Therefore, for large portions of the dynamics the region of poor fitting is avoided. This leads us to believe that the LVC model is sufficient for the aim of this paper, namely to benchmark quantum-classical nonadiabatic dynamics methods against a fully quantum approach to enable future study of large polyene systems. For this purpose, we prefer to use a crude but simple model in order to efficiently scan the parameter space around the values obtained from the fit.

\section{Nonadiabatic Dynamics Methods}\label{sec:methods}
\subsection{Fully Quantum Dynamics via the Short Iterative Lanczos Propagator (SILP) Method}\label{sec:SILP}
Fully quantum dynamics can be approached directly from the quantum-phonon Hamiltonian,~\cref{eq:h_quibronic}. Given a set of electronic states $\{\ket{i}\}$ and a set of vibrational states $\{\ket{v_{\alpha}}\}$ for each mode $\alpha$, we create a basis of product states
\begin{equation}\label{eq:basis_states_silp}
    \ket{n}=\ket{i}\bigotimes_{\alpha=1}^{N_q}\ket{v_{\alpha}}.
\end{equation}
The tensor product is over the number of modes, $N_q$, and $n$ is a composite index over the electronic and vibrational degrees of freedom. For each mode the value of $v_{\alpha}$ lies in the range $0\leq v_{\alpha}\leq D_{\alpha}-1$, where $D_{\alpha}$ is the number of vibrational states in mode $\alpha$. In this basis, the time-evolving wavefunction is

\begin{equation}\label{eq:qph_wfn}
    \ket{\Psi(t)}=\sum_n^{N_H}a_n(t)\ket{n}
\end{equation}
and the starting state is
\begin{equation}
    \ket{\Psi(0)}=\ket{1}\bigotimes_{\alpha=1}^{N_q}\ket{0}.
\end{equation}
This corresponds to the vertically excited $1B_u$ state. For $N_e$ electronic states, the size of the Hilbert space is
\begin{equation} 
    N_H=N_e\prod_{\alpha=1}^{N_q} D_{\alpha}.
\end{equation}
For a fixed number of levels per mode, $D_{\alpha}=D$, the Hilbert space scales exponentially in the number of modes $N_q$. 

For a given number of quantum levels on each mode, one can find the wavefunction of the system at a given time with non-stationary state dynamics, by diagonalizing the Hamiltonian. However, the diagonalization of the Hamiltonian in such a large basis is prohibitively expensive. To address this problem, we use the short iterative Lanczos propagator (SILP) method\cite{paigeComputationalVariantsLanczos1972,parkUnitaryQuantumTime1986,domckeConicalIntersectionsElectronic2004}. This method has previously been used in multi-mode vibronic coupling dynamics in small molecules such as ethene and nitrogen dioxide\cite{koppelStrongNonadiabaticEffects1982,hallerVisibleAbsorptionSpectrum1985} and it is described in Appendix \ref{apx:SILP}.

The Hamiltonian and other observables are calculated using matrix multiplication of sparse matrices. Observables are constructed in the original basis $O_{m,n}=\mel{m}{\hat{O}}{n}$ and found by multiplication with the time-evolving wavefunction $\ket{\Psi(t+\Delta t)}$. The primary observable of interest is the diabatic electronic state populations, for which $\hat{O}=|\Tilde{\psi}_i(t)\rangle\langle \Tilde{\psi}_i(t)|$. For a method with quantum phonons, the diabatic state $|\Tilde{\psi}_i\rangle$ is 
\begin{equation}
    |\Tilde{\psi}_i(t)\rangle=\frac{1}{\sqrt{\braket{\psi_i(t)}}}\ket{\psi_i(t)},
\end{equation}
where
\begin{equation}
    \ket{\psi_i(t)}=\ket{i}\braket{i}{\Psi(t)}=\sum_{v_1}^{D_1}\cdots\sum_{v_{N_q}}^{D_{N_q}}
    a_{i,v_1,\cdots,v_{N_q}}(t)
    \ket{i}\ket{v_1}\cdots |v_{N_q}\rangle.
\end{equation}
Using these states, we construct an $N_e\times N_e$ electronic Hamiltonian with matrix elements
\begin{equation}\label{eq:silp_adia}
    H^{\text{elec}}_{i,j}=\mel{\Tilde{\psi}_i}{\hat{H}_{\text{Q-VC}}-\hat{T}_N}{\Tilde{\psi}_j}.
\end{equation}
The diagonal components of this matrix are the diabatic energies and $\hat{T}_N$ is the quantum mechanical nuclear kinetic energy operator. The adiabatic electronic states and their energies are found by diagonalizing this Hamiltonian at a given time. We also find the expectation value of the position operator $\hat{Q}_{\alpha}$, defined in~\cref{eq:Qalpha_defn}.

Another well-established method for fully quantum dynamics is MCTDH. We have confirmed that SILP and MCTDH lead to the same results for the diabatic populations, as shown in Fig.~S5 of the Supporting Information, which demonstrates that the methods are converged. The reason we use SILP is to compute the adiabatic energies more easily than with MCTDH.

\subsection{Multi-Trajectory Ehrenfest}\label{sec:MTE}
Ehrenfest dynamics is the most basic approach to quantum-classical dynamics. It has well-known flaws, such as failing to describe the branching of wavepackets\cite{tully_molecular_1990,stockClassicalDescriptionNonadiabatic2005}. The force on the nuclei is simply the mean-field force from the electronic wavefunction,
\begin{equation}\label{eq:EF_force_UVP}
    f_{\alpha} = -\frac{\partial\mel{\Psi(t)}{\hat{H}_{\text{C-VC}}}{\Psi(t)}}{\partial q_{\alpha}}=-\mel{\Psi(t)}{\frac{\partial{\hat{H}_{\text{C-VC}}}}{\partial q_{\alpha}}}{\Psi(t)}.
\end{equation}
The wavefunction is propagated according to the velocity-Verlet scheme of Appendix E in Ref.~\citenum{runeson_multi-state_2023}.

In multi-trajectory Ehrenfest (MTE) many Ehrenfest simulations are run, each with different starting conditions. Each trajectory starts in the $1B_u$ diabatic state, but with a Wigner distribution of initial positions and momenta\cite{hilleryDistributionFunctionsPhysics1984}. It includes the quantum zero point fluctuations in positions and momentum of the initial state. At zero temperature, assuming the ground-state potential energy surface is harmonic, the positions and momenta have normal distributions with variance $\sigma_{q_{\alpha}}^2=\hbar/2m\omega_{\alpha}$ and $\sigma_{p_{\alpha}}^2=\hbar m\omega_{\alpha}/2$, respectively, where $\omega_{\alpha}$ is the frequency of coordinate $q_{\alpha}$ in the LVC Hamiltonian, $\omega_{\alpha}=\sqrt{K_{\alpha}^{(0)}/m}$.

For MTE, the observables are simply found from the wavefunction coefficients and averaged over all trajectories. We refer to this approach as wavefunction-type observables.

\subsection{Fewest-Switches Surface Hopping}\label{sec:FSSH}
The Ehrenfest approximation is unable to describe wavefunction bifurcation, which leads to inaccurate dynamics when more than one adiabat is populated simultaneously,
as in Tully's avoided crossing model\cite{mannouch_mapping_2023}.
A way to correct this issue while still exploiting the computational speed of quantum-classical dynamics is to use surface-hopping methods. On each trajectory, the nuclei experience only the force from one adiabat at a time. This adiabat is called the active state and denoted with subscript $a$ in the following. The force on the nuclear coordinates is
\begin{equation}
    f_{\alpha} = -\mel{\phi_{a}}{\frac{\partial{\hat{H}_{\text{C-VC}}}}{\partial q_{\alpha}}}{\phi_{a}}.
\end{equation}

In fewest-switches surface hopping, trajectories hop between adiabatic states stochastically\cite{tully_molecular_1990,tullyMixedQuantumClassical1998,jainPedagogicalOverviewFewest2022}. The wavefunction in the adiabatic basis is
\begin{equation}
    \ket{\Psi(t)}=\sum_ac_a\ket{\phi_a}.
\end{equation}
The adiabats are defined by the eigenvalue equation
\begin{equation}
    \hat{H}_{\text{C-VC}}\ket{\phi_a}=V_a^{\text{ad}}\ket{\phi_a},
\end{equation}
where the eigenvalues, $V_a^{\text{ad}}$, are found by diagonalizing the vibronic coupling Hamiltonian,
\begin{equation}
    \mathbf{V}^{\text{ad}}=\mathbf{U}^{\dagger}\mathbf{H}_{\text{C-VC}}\mathbf{U}.
\end{equation}
For a basis of two states, the transition probability of a hop from active state $a$ to adiabat $b$ for a time step $\Delta t$ is 
\begin{equation}\label{eq:fssh_hop_prob}
    \Pi_{a\rightarrow b} = 2\Delta t\,\text{Re}\left(\frac{c_b T_{ab}}{c_a}\right).
\end{equation}
Negative transition probabilities are set to zero, meaning that there are no hops to a state whose population is decreasing. $T_{ab}$ is the time-derivative coupling,
\begin{equation}\label{eq:T_ab_defn}
    T_{ab}=\braket{\phi_a}{\frac{\partial\phi_b}{\partial t}}
    =\sum_{\alpha}d_{ab}^{({\alpha})}\frac{p_{\alpha}}{m},
\end{equation}
where the nonadiabatic coupling vector $d_{ab}^{({\alpha})}$ is found via the off-diagonal Hellmann-Feynman theorem,
\begin{equation}
    d_{ab}^{({\alpha})}=\frac{\mel{\phi_a}{\frac{\partial\hat{H}_{\text{C-VC}}}{\partial q_{\alpha}}}{\phi_b}}{V^{\text{ad}}_b-V^{\text{ad}}_a}.
\end{equation}

In order for each trajectory to conserve energy, momentum must be rescaled after a hop. Following Tully's original suggestion, we rescale momentum in the direction of the nonadiabatic coupling vector, $d_{ab}^{({\alpha})}$\cite{tully_molecular_1990}. If the nuclear kinetic energy along the nonadiabatic coupling vector is less than the energy gap between the adiabats, the momentum is reversed along the nonadiabatic coupling vector and the active state remains unchanged. 

To calculate the time-derivative coupling, $T_{ab}$, we use the matrix logarithm approach described in Refs.~\citenum{loringComputingLogarithmUnitary2014,jainEfficientAugmentedSurface2016,qiuPracticalApproachWave2023}, because it is more stable at trivial crossings:
\begin{equation}
    \mathbf{T}(t+\Delta t/2)=\frac{1}{\Delta t}\log{\left(\mathbf{U^{\dagger}}(t)\mathbf{U}(t+\Delta t)\right)}.
\end{equation}


It is well known that FSSH suffers from an over-coherence problem\cite{subotnikUnderstandingSurfaceHopping2016,crespo-oteroRecentAdvancesPerspectives2018}. That is, a trajectory evolving on an adiabat that becomes uncoupled to the other adiabats will remain in a superposition of adiabatic electronic states. To address this issue, it is standard to apply a decoherence correction that decays or collapses wavefunction coefficients belonging to a state uncoupled from the active state. We choose to use the instantaneous nonadiabaticity threshold (INT) scheme introduced by Runeson, which has been extensively benchmarked for the Tully models and vibronic coupling models\cite{runesonDecoherenceSurfaceHopping2025}. We use the INT scheme for its computational simplicity and because it relies on a single dimensionless parameter that was found to be less system-dependent than the parameters employed in previous approaches\cite{fangImprovementInternalConsistency1999,granucciCriticalAppraisalFewest2007}. This scheme measures whether states are still coupled to the active state via the dimensionless Massey parameter. For an adiabat $b$ and active state $a$, the Massey parameter is
\begin{equation}\label{eq:massey}
    r_b = \frac{\hbar |T_{ab}|}{|V^{\text{ad}}_b-V^{\text{ad}}_a|}.
\end{equation}
Decoherence occurs when this parameter is less than some threshold,
\begin{equation}
    r_b < r_0,
\end{equation}
where values of $r_0$ between $0.001$ and $0.01$ have been found adequate for a variety of Hamiltonians\cite{runesonDecoherenceSurfaceHopping2025}. In the INT scheme, if this condition is satisfied for state $b$ then its wavefunction coefficient is set to zero, $c_b=0$, and the wavefunction is normalized. We refer to FSSH with this decoherence correction as INT-FSSH.

Observables can be calculated using wavefunction-type observables, but this approach fails to account for frustrated hops. Instead, we find adiabatic populations as the proportion of trajectories with state $a$ as the active state at that time. To get diabatic populations, the rest of the adiabatic density matrix is constructed using $c_a^{\star}c_b$ for coherences. Then, it is rotated to the diabatic basis and the resulting diagonals are the diabatic populations. The decoherence correction ensures that the wavefunction-type and the active state population measures agree qualitatively. These measures disagree strongly without the decoherence correction, as shown in Fig.\ S9 of the Supporting Information.

As for MTE, the normal modes are initiated with a Wigner distribution of positions and momenta and the electronic starting state is the diabat $1B_u$. The probability of starting on an active, adiabatic state $a$ is the modulus squared of the projection of adiabat $a$ onto this starting state.

\subsection{Mapping Approach to Surface Hopping}\label{sec:MASH}
Recently, there have been advances in alternative surface hopping approaches. Taking inspiration from spin-mapping approaches, Mannouch and Richardson introduced the mapping approach to surface hopping (MASH)\cite{mannouch_mapping_2023}. This approach only applies to two states, but Runeson and Manolopoulos first adapted it to multiple states\cite{runeson_multi-state_2023}. Later, Lawrence, Mannouch and Richardson introduced a size-consistent multi-state approach\cite{lawrence_size-consistent_2024}. These methods are summarized in a recent review \cite{richardsonNonadiabaticDynamicsMapping2025}. In the present paper, we use the multi-state MASH method by Runeson and Manolopoulos, because it is considerably simpler than the size-consistent approach.

In MASH, hops between surfaces are deterministic. That is, each trajectory's active state is simply the state of greatest population at any point in time and the force on the nuclei is calculated from the active state, in the same way as FSSH. The MASH direction for momentum rescaling is chosen so that when a frustrated hop occurs and the momentum is reversed, the trajectory moves away from the point of equal populations. This avoids multiple fast crossings. Simply using the nonadiabatic coupling vector $d_{ab}^{({\alpha})}$ as in FSSH does not fulfill this condition. In practice, one derives the time derivative of the population difference between the two adiabatic states in question,
\begin{equation}
    \dot{P}_a - \dot{P}_b = 2\sum_{\alpha}\frac{p_{\alpha}}{m_{\alpha}}\sum_{a'}\text{Re}\left(c_{a^{\prime}}^* d_{a^{\prime}a}^{({\alpha})} c_{a} - c_{a^{\prime}}^* d_{a^{\prime}b}^{({\alpha})} c_{b}\right),
\end{equation}
which defines the direction of momentum rescaling, 
\begin{equation}
    \delta^{({\alpha})}_{ab} = \frac{2}{m_{\alpha}}\sum_{a'}\text{Re}\left(c_{a^{\prime}}^* d_{a^{\prime}a}^{({\alpha})} c_{a} - c_{a^{\prime}}^* d_{a^{\prime}b}^{({\alpha})} c_{b}\right).
\end{equation}

MASH populations calculated using wavefunction-type observables give the incorrect adiabatic and diabatic populations at long time. Therefore, a population estimator is used to treat all bases equally,
\begin{equation}\label{eq:mash_popest}
    \Phi_i=\alpha_{N_e}P_i+\beta_{N_e},
\end{equation}
where for $N_e$ electronic states,
\begin{equation}
    \alpha_{N_e} = \frac{N_e-1}{(\sum_{n=1}^{N_e} 1/n)-1}
\end{equation}
and
\begin{equation}
    \beta_{N_e} = \frac{1-\alpha_{N_e}}{N_e}.
\end{equation}
Using the population estimator ensures correct adiabatic and diabatic populations at thermal equilibrium and treats coherences and populations on the same footing\cite{runeson_multi-state_2023}.

While there are multiple methods to determine the initial wavefunction when using MASH\cite{mannouch_toward_2024}, we choose the simplest method that is consistent with the $\Phi_i$ choice of observable. To start in a diabat $i$, we initialize the wavefunction such that $\Phi_i=1$ and $\Phi_{j\neq i}=0$. These constraints determine the magnitudes $|c_i|$ and we sample the phases uniformly between $0$ and $2\pi$. Just as in the other quantum-classical methods, we start trajectories in a Wigner distribution of position and momentum that is independent of the electronic sampling.

\section{Results}\label{sec:results}
In this Section, we describe how the diabatic populations of our hexatriene LVC model evolved with the quantum-phonon SILP method compare to previous theoretical work. Then, we compare three quantum-classical methods -- MTE, INT-FSSH and MASH -- to our fully quantum benchmark calculations, considering diabatic populations and coordinate displacements. Finally, we analyze where these methods differ in the parameter space around our hexatriene LVC Hamiltonian, varying the vertical energy gap, inter-state coupling constant, intra-state coupling constant and force constant in turn.

\subsection{Hexatriene}\label{sec:hexatriene}

\begin{figure}
    \centering
    \includegraphics[]{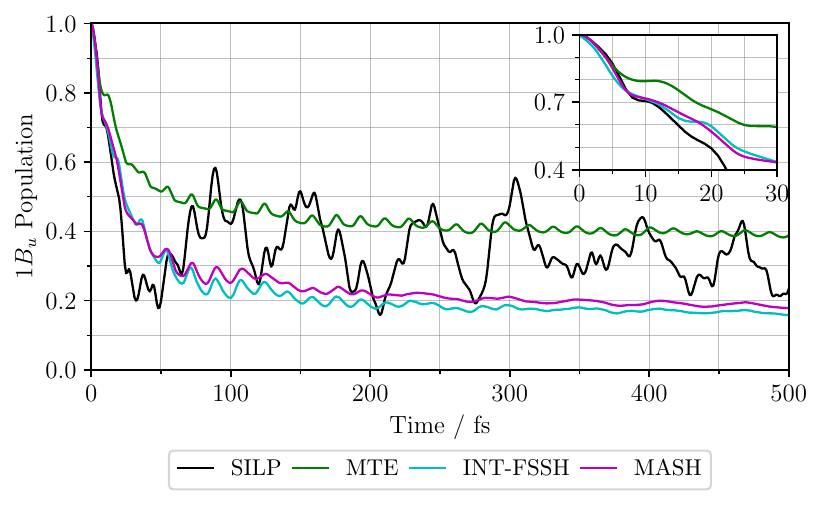}
    \caption{$1B_u$ diabat population plotted against time for four different nonadiabatic dynamics methods. The three quantum-classical methods (MTE, INT-FSSH, MASH) fail to reproduce the quantum-phonon (SILP) results. MTE overestimates the centre of the SILP oscillations, whereas the surface hopping methods underestimate it. The surface hopping methods describe initial dynamics through the avoided crossing slightly more accurately (see inset).}
    \label{fig:hex_2state_pop_comparison}
\end{figure}

Considering internal conversion in \textit{trans}-hexatriene, we want to understand which states are populated and at what timescales. \cref{fig:hex_2state_pop_comparison} shows the population of the $1B_u$ state against time for the four nonadiabatic dynamics methods described in the previous Section. The quantum-phonon Hamiltonian evolved with SILP has a fast decrease in $1B_u$ population to about $0.2$ in $30~\unit{fs}$. Subsequently, there are oscillations with a period of $60$ to $80~\unit{\fs}$ and an amplitude of up to $0.15$, around a population of $0.34$. In previous work, Komainda and coworkers constructed a model with $6$ different in-plane modes and potential energy surfaces calculated with MSCASPT2 and DFT/MRCI\cite{komaindaInitioBenchmarkStudy2016}. Using MCTDH, they found the long-time $1B_u$ population was $0.2$ with MSCASPT2 and $0.4$ with DFT/MRCI. Our results using the extended Hubbard-Peierls Hamiltonian lie within this range, so our model is sufficient to benchmark quantum-classical methods to study polyenes.

\begin{figure}
    \centering
    \includegraphics[]{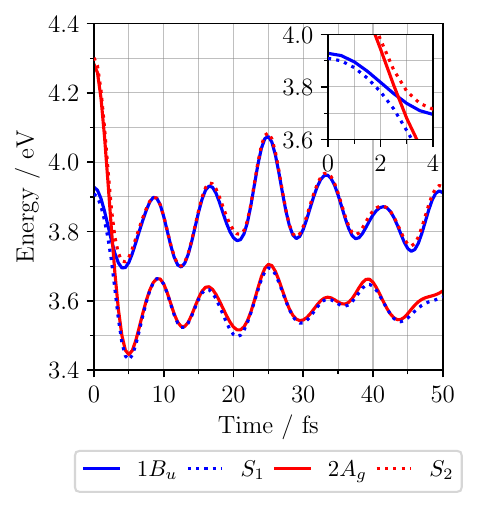}
    \caption{Adiabatic (dotted) and diabatic (solid) potential energies against time for the two-state hexatriene LVC Hamiltonian, calculated with the quantum-phonon SILP method. The diabats cross between $2$ and $3~\unit{\fs}$, while the adiabats show an avoided crossing. This is due to a fast increase in displacement on the optical mode, $Q_5$. The oscillations have a period of about $8~\unit{\fs}$, which is half that of the optical mode.}
    \label{fig:hex_state_adia_energy}
\end{figure}

\begin{figure}
    \centering
    \includegraphics[]{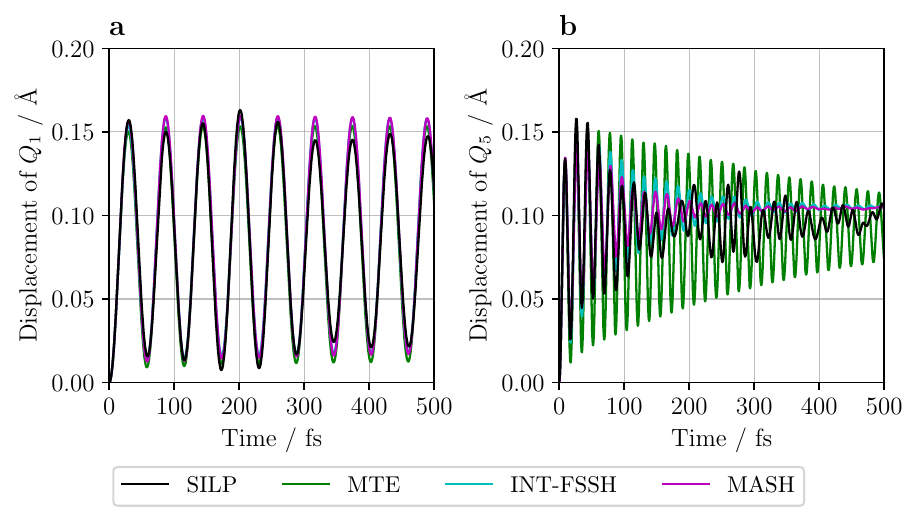}
    \caption{Average coordinate positions against time for the two-state LVC model. (a) The lowest energy mode $Q_1$ oscillates with a time period around $60~\unit{\fs}$. (b) The optical mode $Q_5$ relaxes to about $0.1~\unit{\angstrom}$, which is the minimum of the $2A_g$ potential energy surface. Its faster oscillations with a period of $16~\unit{\fs}$ are the result of its steeper potential energy surface.}
    \label{fig:hex_2state_q_comparison}
\end{figure}

\cref{fig:hex_state_adia_energy} shows the adiabatic and diabatic potential energies of the two-state hexatriene LVC Hamiltonian plotted against time, relative to the energy of the relaxed ground-state diabat and subtracting half the zero point energy. The adiabatic energies, $S_1$ and $S_2$, are found with the quantum-phonon SILP method by diagonalizing~\cref{eq:silp_adia}. Between $2$ and $3~\unit{fs}$ the diabats cross in energy, while the adiabats have an avoided crossing. This is due to a fast increase in displacement on the optical $Q_5$ mode, illustrated in~\cref{fig:hex_2state_q_comparison} (b). After the crossing, $S_1$ and $2A_g$ have the same energy, as do $S_2$ and $1B_u$. Oscillations in these energies occur at a period of about $8~\unit{\fs}$, half that of the optical mode $Q_5$.

Also shown in~\cref{fig:hex_2state_pop_comparison} are the quantum-classical results for MTE, decoherence corrected FSSH (INT-FSSH) and MASH. The MTE and MASH simulations were run with $10^6$ trajectories with a time step of $0.05~\unit{\fs}$. The INT-FSSH simulations were run with $10^5$ trajectories with a time step of $0.005~\unit{\fs}$, because a time step of $0.05~\unit{\fs}$ led to deviations in populations of up to $7\%$ at $25~\unit{\fs}$.  Convergence checks for all methods are presented in the Supporting Information. The two surface hopping methods show a decrease in $1B_u$ population to about $0.65$ after $20~\unit{\fs}$, beyond which they diverge from the quantum-phonon result. The surface hopping approaches underestimate the long-time population as $0.19$ and $0.21$ for INT-FSSH and MASH, respectively. Conversely, MTE overestimates the long-time population to be $0.41$. MTE suffers from overheating, making the $1B_u$ population closer to $0.5$ than it would be otherwise. 

None of the quantum-classical methods studied are able to reproduce the complex  $1B_u$ population oscillations of the quantum-phonon SILP method. The population obtained with MTE oscillates with a time period around $16~\unit{fs}$, similar to the oscillations on the mode $Q_5$, shown in~\cref{fig:hex_2state_q_comparison} (b). The surface hopping approaches also have oscillations of $16~\unit{fs}$ until $200~\unit{\fs}$, when, as~\cref{fig:hex_2state_q_comparison} (b) shows, oscillations on this mode decay in time. 

Next, we compare nuclear positions against time for the quantum-phonon SILP method and the quantum-classical methods.~\cref{fig:hex_2state_q_comparison} (b) shows the average classical nuclear positions against time for the highest energy mode, $Q_5$, which oscillates with a period of $16~\unit{\fs}$. For SILP and MTE, the optical mode $Q_5$ relaxes to about $0.95~\unit{\angstrom}$ which, as shown in~\cref{fig:PES_Q5}, is close to the minimum of the $2A_g$ PES. For the surface hopping methods, the $Q_5$ mode relaxes to $1.05~\unit{\angstrom}$. This is closer to the minimum of the $2A_g$ PES, because of the larger $2A_g$ population for these methods. The oscillations almost vanish for the surface hopping methods, while they remain with some amplitude for MTE and SILP. As shown in~\cref{fig:hex_2state_q_comparison} (a), the lowest energy symmetric mode, $Q_1$, oscillates at an amplitude of $0.08~\unit{\angstrom}$ and a period of $58~\unit{\fs}$. The longer time oscillations are due to a lower curvature PES and the quantum-classical methods successfully reproduce these oscillations. Interestingly, the $Q_1$ period is comparable to the timescale of the population oscillations. However, the latter oscillations persist even if the $Q_1$ mode is excluded from the model (as shown in  Fig.\ S10 of the Supplementary Information), which indicates a more complicated vibronic interference pattern that the quantum-classical methods used here are unable to fully describe. For the anti-symmetric modes $Q_2$ and $Q_4$, the expectation value of the displacement is zero.

\subsection{Parameters Scan}
\begin{figure}
    \centering
    \includegraphics[]{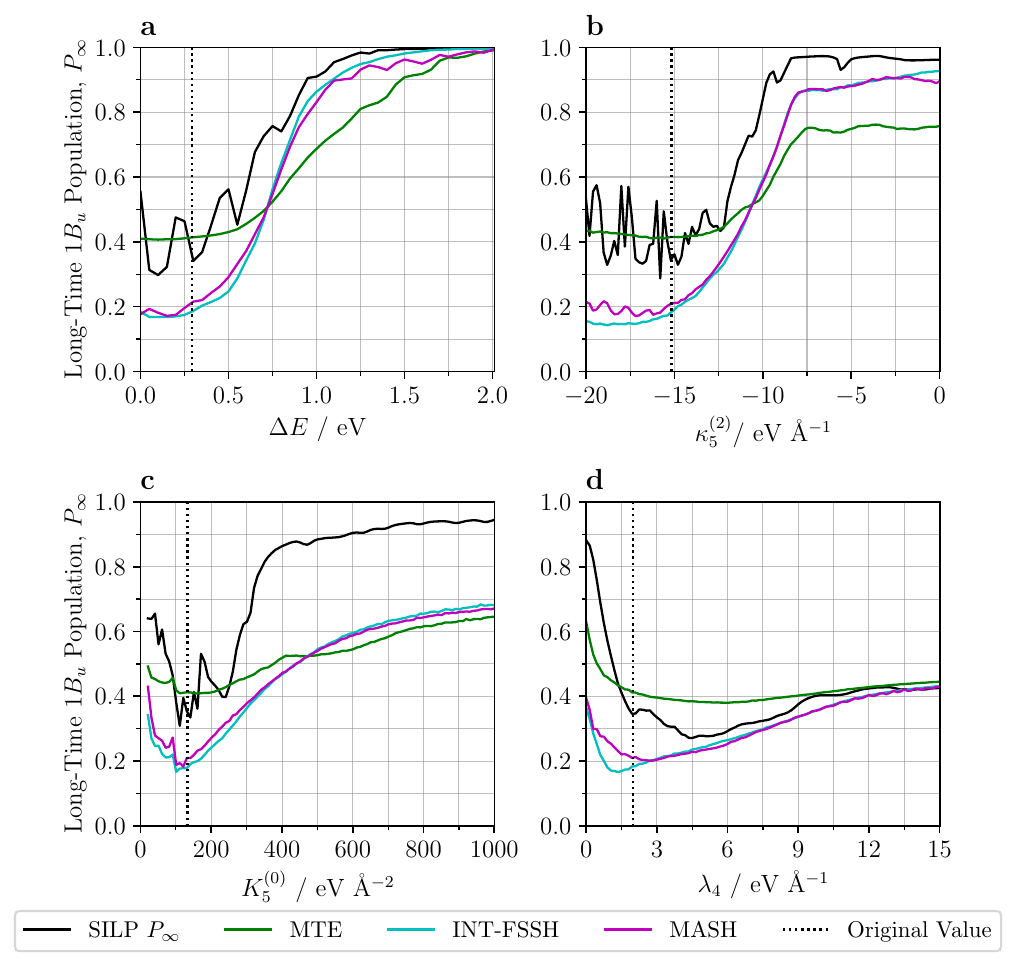}
    \caption{The long-time $1B_u$ population, $P_{\infty}$, found from fitting to an exponential decay is plotted against various parameters of the hexatriene LVC Hamiltonian. The vertical black dotted lines indicate the physical hexatriene parameter set given in Table 1. (a) There is a decrease in internal conversion as the energy gap increases. (b) Increasing the intra-state coupling constant results in reduced population transfer from the $1B_u$ because the $2A_g$ minimum rises above the $1B_u$ minimum. (c) Increasing the force constant results in lower internal conversion because of a higher energy crossing point. Decreasing the force constant also reduces internal conversion because the crossing point is far from the $1B_u$ minimum. (d) Approaching zero inter-state coupling, $P_{\infty}$ increases. For large coupling, the long-time population tends to $0.5$.}
    \label{fig:param_scan_2x2_pop_inf}
\end{figure}

To decide upon the most reliable quantum-classical method for more general polyene systems, we scan the parameter space around the hexatriene parameter set. We vary four parameters in turn: the vertical energy gap, $\Delta E=E^{(2)}-E^{(1)}$; the intra-state coupling constant of the highest energy mode on $2A_g$, $\kappa_5^{(2)}$; the force constant on mode $Q_5$ shared by all states, $K_5^{(0)}$; and the inter-state coupling constant on the highest energy anti-symmetric mode $Q_4$, $\lambda_4$. We compare the extent of internal conversion, as measured by the long-time population. For each simulation, we find the long-time $1B_u$ population by fitting to an exponential decay,
\begin{equation}\label{eq:param_scan_fit}
    P_{1B_u}(t)\approx (1-P_{\infty})e^{-t/\tau}+P_{\infty},
\end{equation}
using least squares regression. The extracted long-time $1B_u$ population, $P_{\infty}$, is shown in~\cref{fig:param_scan_2x2_pop_inf} for a range of parameters. The hexatriene parameter values are shown by vertical dotted lines.

The standard deviation of $P_{\infty}$ is shown in Fig.~S11 of the Supporting Information. The standard deviation of $P_{\infty}$ is always less than $0.0075$ for all methods, except a single anomalous data point that arises from a poor exponential fit and was removed. SILP has an order of magnitude larger standard deviation than the other methods, due to its larger amplitude oscillations coming from its quantum treatment of the nuclei. The sometimes rapid oscillations in the SILP $P_{\infty}$ values reflect strong sensitivity to the Hamiltonian parameters, not poor fitting. This is supported by Fig.~S12 of the Supporting Information, which shows that $P_{\infty}$ gives the same result as the long-time mean of populations for several different time ranges.

\cref{fig:param_scan_2x2_pop_inf} (a) shows how the energy gap between the excited states affects the long-time population of $1B_u$. As the vertical energy gap increases, $P_{\infty}$ increases. As expected, a larger energy gap suppresses internal conversion. Around the physical parameter regime, MTE gets similar long-time populations to SILP, but it does not reproduce the oscillations in $P_{\infty}$ seen in the SILP calculations. We attribute these oscillations to resonance between the energy gap and the highest energy anti-symmetric mode $Q_4$, which has frequency $0.23~\unit{\eV}$. For an energy gap greater than $0.75~\unit{\eV}$, surface hopping recovers the long-time population more accurately than MTE. Approaching an energy gap of $2.0~\unit{\eV}$, dynamics is effectively adiabatic and MTE, INT-FSSH and MASH reproduce the quantum-phonon result of only the $1B_u$ state being populated.

Next, we consider how the intra-state coupling constant, $\kappa_5^{(2)}$, affects the long-time population. As shown in~\cref{fig:param_scan_2x2_pop_inf} (b), for large negative couplings, MTE more accurately reproduces the quantum-phonon result than the surface hopping approaches do. However, the quantum-phonon result has large fluctuations absent from the MTE result, which we attribute to resonances in the quantum levels of the two potential energy surfaces. For all methods, as the magnitude of $\kappa_5^{(2)}$ decreases, $P_{\infty}$ increases. For the quantum-phonon method, the population of $1B_u$ is close to unity near $\kappa_5^{(2)}=0$, because the $2A_g$ minimum rises above the $1B_u$ minimum. For values of $\kappa_5^{(2)}$ greater than $-12~\unit{\eV\angstrom^{-1}}$, no quantum-classical method gets quantitative agreement with the quantum-phonon result. However, INT-FSSH and MASH better reproduce the shape of the curve, but with populations between $0.1$ and $0.2$ too low.

\cref{fig:param_scan_2x2_pop_inf} (c) shows that as the force constant increases, so does the long-time population of $1B_u$. As the potential energy surfaces become steeper, the crossover with the $1B_u$ increases in energy, reducing the extent of internal conversion. For large force constants, no quantum-classical method reproduces the quantum-phonon result. For force constants closer to the hexatriene value, MTE best reproduces the SILP result. In the small force constant regime, the flattened potential energy surfaces reduce internal conversion by displacing the crossing point far beyond the $1B_u$ minimum. The surface hopping methods better describe the overall trend, but they severely overestimate the degree of internal conversion.

Considering the inter-state coupling constant, which is varied in~\cref{fig:param_scan_2x2_pop_inf} (d), all methods show similar trends at small and large values of $\lambda_4$. For very small values, the coupling of the two states is turned off and the long-time $1B_u$ population is large, because it is the starting state. $P_{\infty}$ does not tend to unity as $\lambda_4$ approaches zero, because the inter-state coupling on the other anti-symmetric mode, $\lambda_2$, is non-zero. While SILP predicts almost no internal conversion when $\lambda_4=0$, all quantum-classical methods predict a larger degree of internal conversion, perhaps due to higher amplitude oscillations on the $Q_4$ mode. For large values, the coupling mixes the states strongly, so $P_{\infty}$ tends to $0.5$. In between, all methods have a minimum, but no quantum-classical method replicates the position or depth of the minimum of the quantum-phonon result.

In the preceding discussion MASH and INT-FSSH were often referred to collectively as the surface hopping approaches. This is because their long-time $1B_u$ populations are remarkably similar across a wide range of parameters, despite their different hopping criteria, observables, initial electronic distributions and decoherence treatments.

\section{Conclusions}\label{sec:conclusion}
This paper has described a protocol for performing nonadiabatic excited state dynamics in polyenes, with the aim of applying it to carotenoids with $C_{2h}$ symmetry. To that end, we have taken hexatriene as a model system and constructed a linear vibronic coupling model from the one-dimensional extended Hubbard-Peierls Hamiltonian. Using exact diagonalization we calculated the potential energy surfaces and the inter-state coupling constants of low-lying electronic states.

Following photoexcitation, we simulated the nonadiabatic dynamics for the quantum-phonon LVC Hamiltonian, using the short iterative Lanczos propagator (SILP) method. The long-time $1B_u$ population of the model gave results within the range of previous theoretical results found using \textit{ab initio} electronic structure methods and MCTDH\cite{komaindaInitioBenchmarkStudy2016}, indicating the robustness of our protocol.

Using SILP results as a benchmark, we assessed the performance of three quantum-classical approaches, which are able to more efficiently simulate excited state dynamics in polyenes. We compared performance for the two-state LVC model at the physically relevant parameter set and for parameter scans about these values. 

For the two-state hexatriene LVC model, multi-trajectory Ehrenfest overestimated the long-time population while the surface hopping methods underestimated the long time population. For ultrafast times (up to $15~\unit{\fs}$) the surface hopping approaches gave the better agreement with quantum-phonon populations. All methods gave correct long-time displacements of the normal coordinate $Q_1$, while the surface hopping approaches reproduced the quantum-phonon simulations for the longest time on $Q_5$. However, no quantum-classical approach correctly reproduced the oscillations of the populations.

We also varied the parameters of the LVC Hamiltonian and compared the fitted long-time $1B_u$ populations. MTE gave good agreement with the SILP benchmark around the hexatriene parameter regime for the energy gap, force constant and intra-state coupling constant scans, but it did not reproduce the quantum oscillations along these parameters. INT-FSSH and MASH gave remarkably similar long-time populations, despite large differences in the two approaches. The surface hopping approaches successfully reproduced the trends in all scans, although they consistently over-predicted the degree of internal conversion.

In future work, we plan to apply the methodology described in this paper to develop a vibronic coupling model for carotenoids (in particular, lycopene and zeaxanthin). As mentioned in the Introduction, this will incorporate three excited electronic states and we will use DMRG to compute the excited states of the extended Hubbard-Peierls Hamiltonian. The LVC model has limitations, particularly at dissociation and at long times. We also anticipate that  carotenoid energy surfaces will be anharmonic, so we will go beyond the linear term in the vibronic coupling model for carotenoids. We will use surface hopping approaches to simulate the ultrafast dynamics of the vibronic coupling Hamiltonian in which all the normal modes are included. We expect that the fully quantum population oscillations will decay more rapidly for these larger molecules, as is generally observed when including more nuclear modes\cite{komaindaInitioBenchmarkStudy2016}, which justifies the use of quantum-classical methods. However, we will also perform fully quantum simulations on a reduced-dimensionality vibronic coupling model in which only the key normal modes are incorporated, to improve our confidence in our prediction of long-time populations. For example, these will be the high-energy symmetric modes that cause the diabatic energy crossovers and the low-energy anti-symmetric modes that drive the internal conversion between the $B_u$ and $A_g$ electronic manifolds. 

\begin{acknowledgement}
T.N.G. is grateful for financial support from the Department of Chemistry, University of Oxford, the Engineering and Physical Sciences Research Council (grant number EP/W524311/1) and Balliol College, University of Oxford for a Balliol Scholarship and Jowett Scholarship. T.N.G. thanks Joseph Cooper for useful discussions. J.E.R. was funded by a Research Fellowship from the Alexander von Humboldt Foundation. The authors would like to acknowledge the use of the University of Oxford Advanced Research Computing (ARC) facility in carrying out this work. 
\end{acknowledgement}

\begin{suppinfo}
The following files are available free of charge.
\begin{itemize}
  \item Supporting Information: Contains Hamiltonian parameters, cuts of the potential energy surfaces, convergence tests for all dynamics methods, a comparison of FSSH observables, fully quantum population dynamics omitting the $Q_1$ mode and extra details of the parameter scan.
\end{itemize}

\end{suppinfo}

\bibliography{main.bib}

\appendix
\section{The Short Iterative Lanczos Propagator Method}\label{apx:SILP}
This Section explains in more detail the short iterative Lanczos propagator (SILP) method introduced in~\cref{sec:SILP}. The SILP method reduces the full Hilbert space to a much smaller Krylov subspace of size $N_L\ll N_H$. The Krylov space is found by recursively multiplying an initial state, $\ket{k_0}=\ket{\Psi(t)}$, by the Hamiltonian as follows:
\begin{equation}\label{eq:l1_defn}
    \ket{l_1}=\hat{H}_{\text{Q-VC}}\ket{k_0}-\mel{k_0}{\hat{H}_{\text{Q-VC}}}{k_0}\ket{k_0},
\end{equation}
\begin{equation}\label{eq:k1_defn}
    \ket{k_1}=\frac{1}{\sqrt{\braket{l_1}}}\ket{l_1}.
\end{equation}
The vector $\ket{l_1}$ is the result of the Hamiltonian applied to the initial state, subtracting the component on the initial state. It is then normalized to give the next state in the Krylov subspace, $\ket{k_1}$. The rest of the subspace is found via Lanczos iteration
\begin{equation}\label{eq:lj+1_defn}
    \ket{l_{\nu+1}}=\hat{H}_{\text{Q-VC}}\ket{k_{\nu}}-\mel{k_{\nu}}{\hat{H}_{\text{Q-VC}}}{k_{\nu}}\ket{k_{\nu}}-\sqrt{\braket{l_{\nu}}}\ket{k_{{\nu}-1}},
\end{equation}
\begin{equation}\label{eq:kj+1_defn}
    \ket{k_{{\nu}+1}}=\frac{1}{\sqrt{\braket{l_{{\nu}+1}}}}\ket{l_{{\nu}+1}}.
\end{equation}
The Krylov vectors $\ket{k_{\nu}}$ give the columns of the $N_H\times N_L$ matrix $\mathbf{K}$ that transforms the Hamiltonian into a tridiagonal matrix $\mathbf{X}$,
\begin{equation}
    \mathbf{X=K^{\dagger}HK}.
\end{equation}
The matrix elements of $\mathbf{K}$ are $K_{n,{\nu}}=\braket{n}{k_{\nu}}$. While calculating the Lanczos iterations, we also find the matrix elements of $\mathbf{X}$:
\begin{equation}
    X_{{\nu},{\nu}}=\mel{k_{\nu}}{\hat{H}_{\text{Q-VC}}}{k_{\nu}} 
\end{equation}
\begin{equation}
    X_{{\nu},{\nu}+1}=X_{{\nu}+1,{\nu}}=\sqrt{\braket{l_{{\nu}+1}}} 
\end{equation}
\begin{equation}
    X_{{\nu},\mu} = 0 \text{ for } |\nu-\mu|>1.
\end{equation}

Now that we have the Hamiltonian reduced to tridiagonal form, we diagonalize it to perform dynamics,
\begin{equation}
    \mathbf{Y=S^{\dagger}XS}.
\end{equation}
A vector $\ket{q}$ in the Krylov basis evolves as
\begin{align}
    \ket{q(t+\Delta t)} &= \exp(-i\mathbf{X}\Delta t/\hbar)\ket{q(t)} \\
    &= \mathbf{S}\exp(-i\mathbf{Y}\Delta t/\hbar)\mathbf{S^{\dagger}}\ket{q(t)} \\
    &= \mathbf{P}\ket{q(t)}.
\end{align}
Since our initial state is always the zeroth Krylov vector, $\ket{q(0)}=\ket{k_0}$, we only need the first column of $\mathbf{P}$ to get the time evolving wavefunction,
\begin{equation}
    \ket{q(t+\Delta t)} = \sum_{{\nu}=0}^{N_L-1}P_{{\nu},0}\ket{k_{\nu}}.
\end{equation}
In the original product basis, we find
\begin{align}
    \ket{\Psi(t+\Delta t)}=\sum_n^{N_H}\sum_{{\nu}=0}^{N_L-1}K_{n,{\nu}}P_{{\nu},0}\ket{n}.
\end{align}
In the limit $N_L$ approaches $N_H$ this method is exact, but remarkably as few as $10$ Krylov vectors are required for accurate results for a limited time, defined below. 

After a certain amount of time, the wavefunction will evolve out of the Krylov subspace. To see when this is happening, we set a threshold, $\epsilon$, on the probability that the wavefunction exists in the last Krylov vector, $\ket{k_{N_L-1}}$. For a given $N_L$, initial Krylov vector $\ket{k_0}$ and tolerance $\epsilon$, the maximum amount of time that the Krylov space can be accurately evolved for is approximately\cite{parkUnitaryQuantumTime1986}
\begin{equation}
    \tau_{\epsilon} \approx \left(\epsilon\left(\frac{(N_L-1)!}{\prod^{N_L-2}_{\nu=0}X_{\nu,\nu+1}/\hbar}\right)^2\right)^{1/2(N_L-1)}.
\end{equation}
At this point, we calculate a new Krylov subspace where the initial vector is $\ket{k_0}=\ket{\Psi(\tau_{\epsilon})}$. In doing so, we update $\mathbf{K, X, Y, S}$ and $\mathbf{P}$, then find $\tau_{\epsilon}$ for the new subspace. We repeat this process until the total time is reached.

\end{document}